# Comprehensive Analysis Method of Acquiring Wall Heat Fluxes in Rotating Detonation Combustors

Yingchen Shi, Yongbo Zhang, Haocheng Wen[*], Bing Wang[*]

School of Aerospace Engineering, Tsinghua University, Beijing 100084, China.

[*]Corresponding author: haochengwenson@126.com and wbing@tsinghua.edu.cn.

**Abstract:** Accurate perception of the combustor thermal environment is crucial for thermal protection design of a rotating detonation combustor (RDC). In this study, a comprehensive analysis method is established to calculate the non-uniform heat flux distribution of the RDC by utilizing the measured temperature distribution of the combustor outer wall obtained by the high-speed infrared thermal imager. Firstly, in order to determine the inverse heat flux solving method, a physical model based on the geometric characteristics of the RDC is constructed and its thermal conductive process is simulated, given by different heat flux boundary conditions. Then the wall heat fluxes are inversely calculated by the Levenberg-Marquardt (L-M) method based on the above numerical data. Results show that the L-M method can obtain more accurate heat flux distribution even in the zones with large heat flux gradients, considering the axial heat conduction within the combustor outer wall caused by the non-uniform heat flux. Finally, the wall heat flux distribution is analyzed coupling the L-M method together with the experimental measurements in kerosene two-phase RDC. The analyses show that the highest temperature of combustor outer surface and the highest wall heat flux occurs within the region of 20mm from the combustor head, which corresponds to nearly 14% of the combustor length. With the increase of axial distance, the heat flux is rapidly reduced, and then the heat flux distribution is more uniform at the downstream region of the combustor. The heat flux peak and thermal heat rate are positively correlated with the combustor equivalence ratio in the range between 0.44 and 0.64.

**Keywords:** rotating detonation combustor, heat flux, analysis method, Levenberg-Marquardt method,

## Nomenclature

| | | | |
|---|---|---|---|
| $c$ | heat capacity of combustor outer wall [J/(kg·K)] | $T_\infty$ | environment temperature [K] |
| $h$ | convective heat transfer coefficient between outer wall and environment [W/(m²·K)] | $T$ | temperature distribution in the physical model [K] |
| $h_c$ | contact thermal conductivity [W/(m²·K)] | $\delta$ | thickness of the combustor outer wall [mm] |
| $L$ | length of the combustor outer wall [mm] | $\rho$ | density of combustor outer wall [kg/m³] |
| $q_{spec}$ | specified heat flux distribution on the inner surface of combustor outer wall [MW/m²] | $\lambda$ | thermal conductivity coefficient of outer wall [W/(m·K)] |
| $q_{cal}^{0D}$ | heat flux calculated by 0D method [MW/m²] | $\dot{m}_O$ | mass flow rate of oxidizer [kg/s] |
| $q_{cal}^{LM}$ | heat flux calculated by L-M method [MW/m²] | $\dot{m}_F$ | mass flow rate of kerosene [kg/s] |
| $T_{wf}$ | temperature of the combustor outer surface [K] | ER | Equivalence Ratio |
| $T_{cal}$ | calculated temperature of the outer surface [K] | MFR | Mass Flow Rate |

# 1. Introduction

Detonation is a supersonic combustion process characterized by high theoretical thermodynamic cycle efficiency and self-pressurization[1,2]. As a highly promising means of propulsion, rotating detonation combustor (RDC) has attracted considerable attention in recent years. Researchers have conducted studies on the propagation stability of rotating detonation waves[3–7] and the formation mechanism of combustion modes in RDC[8–10]. Moreover, experimental prototypes have been developed such as rotating detonation rocket engines[11,12], rotating detonation turbine engines/gas turbines[13,14], and rotating detonation scramjet engines[15–18]. These efforts have resulted in significant progress in the engineering practicality of rotating detonation power systems. Additionally, researchers have preliminarily applied cooling methods such as film cooling[19,20] and thermal protection coatings[21] to RDC. However, the understanding on the thermal environment of the RDC is still insufficient, and researches on thermal protection design and thermal management methods are still rare. Due to these factors, the reliable long-term operation of the RDC is difficult to achieve, which limits the development of rotating detonation power devices.

In order to obtain the authentic thermal environment of RDCs, researchers have conducted a series of experiments to analyze the heat flux distribution and the parametric influence on the distribution. The measurement methods employed in these experiments can be broadly classified into three categories, namely contact measurement methods (such as thermocouples[22–25] and heat flux sensors[26–28]), thermal equilibrium measurement methods (such as the water-cooled jacket[29,30]), and non-contact measurement methods (such as the infrared thermal imager[31]).

In the early 1990s, Bykovskii[22] investigated the discrepancy of heat flux in an annular rotating detonation combustor under detonation and deflagration modes. The heat flux was calculated by the temperature data obtained by the thermocouples installed on the combustor outer wall. The results indicated a substantial reduction in the axial heat flux peak when the RDC was operated in detonation mode, as compared to the deflagration mode, which means that detonation combustion is more favorable for wall cooling. The results also showed that the axial heat flux exhibited a distribution pattern of initially increasing and then decreasing. Subsequently, researchers studied the distribution of heat flux[23] and its influencing factors[24] based on the temperature data from thermocouples. Theuerkauf et al.[26,27] and Meyer et al.[28] measured heat flux waveforms in the RDC by utilizing self-developed high-frequency heat flux sensors. Furthermore, the axial variation of heat flux[27,28] and the influencing factors of heat transfer coefficient[28] were analyzed based on the experimental data. The use of single-point measurement techniques, such as thermocouples and heat flux sensors, enable high-frequency measurement of temperature or heat flux at individual points. However, the limited number of measurement points makes it difficult to obtain the distribution of temperature fields or heat flux fields.

Theuerkauf et al.[29] designed a RDC with a water-cooled jacket and compared the heat rate of the combustor under detonation and deflagration modes. The results indicated that when the mode changes from deflagration to detonation, the heat rate inside the combustor significantly increased. Aliakbari et al.[30] designed a water-cooled jacket arranged circumferentially along the axial direction of the RDC to study the heat flux at different axial positions of the combustor. The results demonstrated that the axial distribution of average heat flux was related to the combustion mode. Additionally, they also found that the heat flux was more sensitive to changes in ER than mass flow rate (MFR). Even though, it is difficult to obtain the high-spatial-resolution heat flux distribution using the thermal equilibrium method, and the geometric constraint of water-cooled jacket makes it difficult to install other sensors.

The use of infrared thermal imager enables the non-contact temperature measurement of the RDC. Zhou et al.[31] investigated the distribution and influencing factors of heat flux in the RDC using thermocouples and infrared thermal imager. The results indicated that the heat flux peak is situated in the same region as the detonation wave propagation zone, and the heat rate of the RDC increased with the rise of ER. This work applied the infrared thermal imager to the measurement of heat flux in the RDC. The applicability and error of the heat flux calculation method based on the infrared thermal imager have not been analyzed yet.

In summary, compared to single-point measurement methods, the infrared thermal imager enables the measurement of temperature field distribution, and can provide more comprehensive information for understanding the thermal environment of RDCs. Nevertheless, the applicability and inaccuracies of existing heat flux calculation method that rely on temperature data obtained by infrared thermal imager have not been discussed. Therefore, an accurate method for calculating heat flux is of significant importance for the thermal protection design of RDCs.

In this study, the experiments are performed on the kerosene two-phase rotating detonation combustion experimental system, and the real-time temperature distribution on the surface of combustor outer wall is obtained using the infrared thermal imager. The applicability of the zero-dimensional (0D) method and the Levenberg-Marquardt (L-M) method in calculating the heat flux distribution within RDC are compared. Furthermore, the heat flux distribution is calculated using the experimental temperature data obtained by the infrared thermal imager, and the influence of equivalence ratio (ER) on the heat flux distribution is discussed. This work provides an accurate method to study the thermal environment of RDCs and serves the thermal protection design of RDCs.

## 2. Experimental methodology

The experimental system of kerosene two-phase rotating detonation used in this study was composed of the gas supply system, fuel supply system, ignition device, data acquisition system, and timing control system[32], as shown in Figure 1. The oxidizer used in the experiments was the mixture of nitrogen and oxygen, with the oxygen volume fraction of 40%, and the fuel used in the experiments was RP-3 kerosene. The oxidizer was stored in a high-pressure storage tank with a volume of 2 m$^3$. The MFR of the oxidizer $\dot{m}_O$ was regulated through an electric pressure regulating valve. The supply of the oxidizer was controlled by a solenoid valve, and the $\dot{m}_O$ was measured by a mass flow meter (EMERSON K200S). The kerosene was stored in a storage tank and was supplied by nitrogen extrusion system. The MFR of kerosene $\dot{m}_F$ was achieved by changing the pressure of the storage tank. The supply of kerosene was controlled by a solenoid valve and $\dot{m}_F$ was measured by a turbine flow meter (MEACON DN10). A high-speed data acquisition instrument (NI cDAQ) was used to collect sensor signals, and the experiment system was controlled by a PLC device.

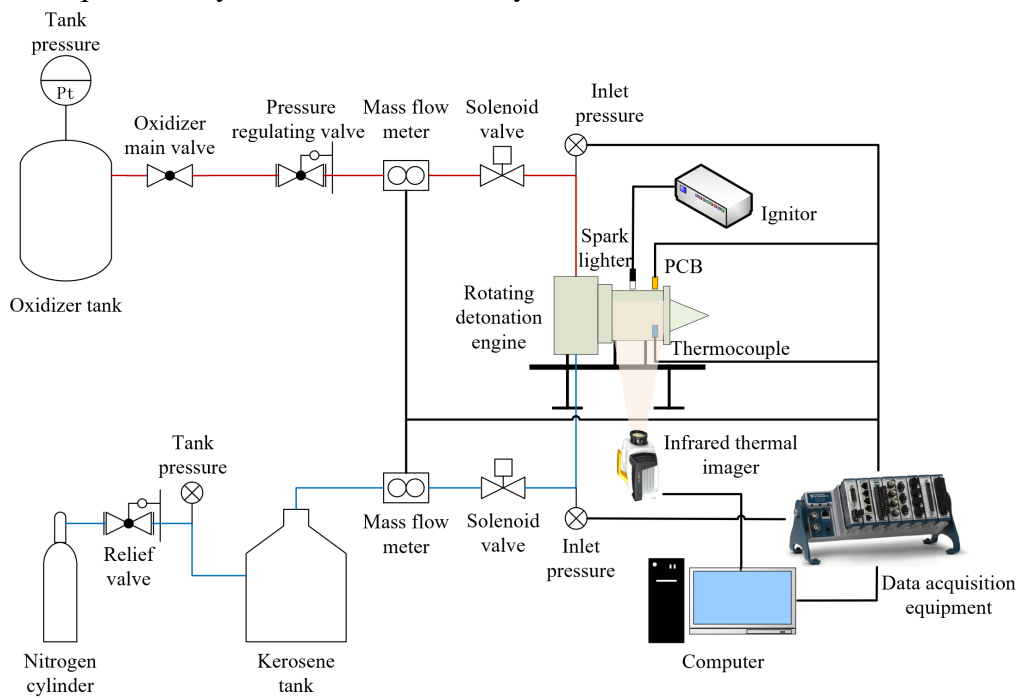

Figure 1. Schematic diagram of two-phase rotating detonation combustion experiment system

In this study, a high-speed infrared thermal imager (TELOPS FAST M200) was used to record the temperature distribution on the surface of the combustor outer wall. The resolution of the filming window was set to 640×512 and the acquisition frequency was set to 50 Hz. In addition, in order to calibrate the temperature data obtained by the thermal imager, a film K-type thermocouple (RKC ST-50, range 0-300°C, accuracy ±1.3°C) was pasted on the temperature measurement area to obtain the temperature at the measurement points. The response time required to track 95% of the instantaneous temperature changes of the measured object is 80 ms for temperature ranging from 150-200°C. A high-frequency pressure sensor (PCB 113B24) was placed 15 mm from the entrance of the combustor to record the high-frequency pressure inside the combustor.

Figure 2 shows a schematic diagram of the kerosene two-phase RDC. The kerosene is atomized and injected through 20 groups of coaxial centrifugal injectors evenly distributed along the circumference of the injection panel. The oxidizer enters the combustor through the annular gaps around the injectors[32]. The outer diameter of the combustor inner wall is 65 mm, the inner diameter of the outer wall is 100 mm, the width of the annular channel is 17.5 mm, and the length of the combustor is 145 mm.

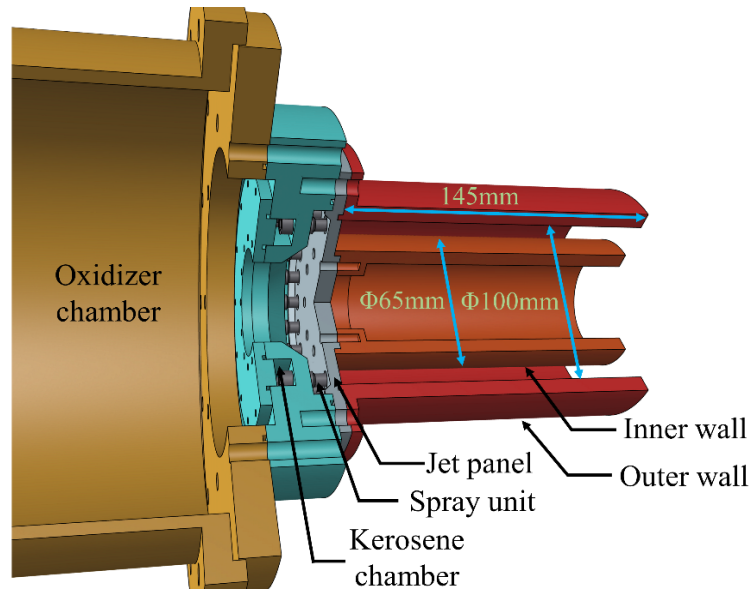

Figure 2. Schematic diagram of the kerosene two-phase RDC

The temperature measurement area was milled on the outer wall of the RDC to reduce the influence of wall heat capacity on heat flux measurement, as shown in Figure 3. The wall thickness of the temperature measurement zone was 3 mm, and the thickness was 8 mm in other areas of the outer wall. The angle of the temperature measurement zone along the circumferential direction of the RDC was 74°. Additionally, in order to improve the emissivity of the temperature measurement zone and improve the accuracy of temperature measurement, a high-temperature resistant paint with an emissivity (0.90) was sprayed on the temperature measurement zone.

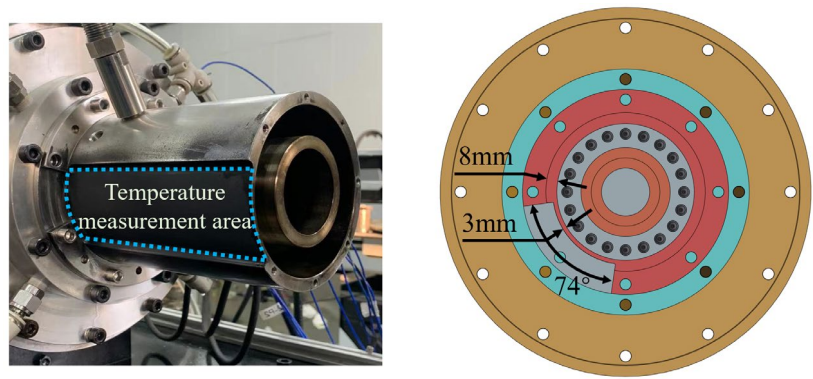

Figure 3. Schematic diagram of temperature measuring area on the outer wall

Figure 4 illustrates the time sequence of the experiments. The operation duration of the combustion device was set at 1000 ms, to acquire sufficient valid data of the infrared thermal imager and protect the high-frequency pressure sensor from damage.

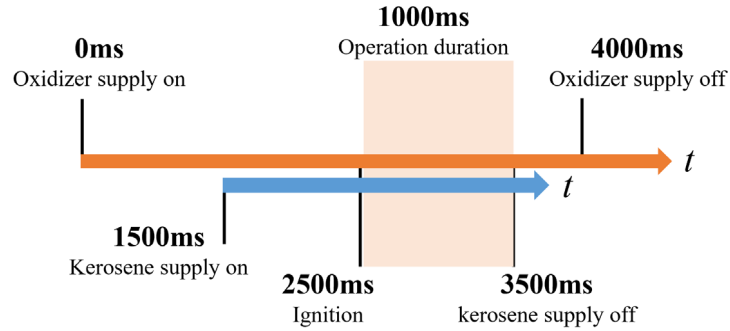

Figure 4. Schematic diagram of the experimental sequence

## 3. Identification of inverse calculation methods

This section firstly provides the definition of a physical model and the research process of inversely solving the heat flux. Then the heat flux calculation methods that inversely solve the heat flux distribution within RDC based on the temperature of the combustor outer surface are introduced. Finally, the error and the applicability of the calculation methods are compared and discussed.

### 3.1 Inverse problem of solving heat fluxes

The physical model used in this study and the research process of inversely solving the heat flux are presented as follows. Before delving into these aspects, it is necessary to state some assumptions:

1): The RDC exhibits high-frequency variation in the flow field, often exceeding several thousand hertz. However, this variation does not have significant impact on the temperature of the combustor outer surface due to the thermal capacity of the outer wall. Therefore, when studying the heat conduction of the thermal fluid within RDC to the outer wall, the high-frequency periodic variation of the flow field is ignored. The relevant physical quantities in the RDC are constantly invariant when the injection conditions and combustion mode remains const.

2): Due to the short working time of the RDC and the limited increase of the outer wall temperature, the changes in heat flux distribution and the temporal evolution of the material properties of the combustor outer wall are disregarded in this study.

3): The temperature measurement results reveals that the temperature distribution of the outer wall exhibits non-uniformity in the circumferential direction accompanied by period changes, which indicates the presence of circumferential heat conduction on the outer wall. Nonetheless, in comparison to the axial temperature gradient of the outer wall, the circumferential temperature gradient is negligible. Consequently, the circumferential heat conduction within the outer wall is neglected in this study, in order to focus on studying the heat flux distribution within RDC along the axial direction, which is expressed as $q_{spec}$.

The physical model investigated in this study is shown in Figure 5. The radial section of the combustor outer wall is selected as the physical model, and the length in *x* and *y* directions corresponds to the axial length and wall thickness of the outer wall, respectively. Note that the influence of the boundary conditions on the temperature distribution in the surrounding area should be considered while calculating the heat flux based on the temperature of the combustor outer surface. However, existing experimental studies that solved the heat flux within RDC based on the temperature of the combustor outer surface have neglected the influence of boundary conditions, particularly the heat conduction between the head of the combustor outer wall and other structures. In this study, the impact of boundary heat conduction is manifested in the brown area at the top of the physical model, as shown in Figure 5, where heat flux is zero. This area corresponds to the locating slot of the outer wall, which serves as a critical component for ensuring the accurate installation and sealing of the RDC. The origin point of the *y* coordinate corresponds to the inner surface of the outer wall, and the origin point of the *x* coordinate corresponds to the exit of the injection panel.

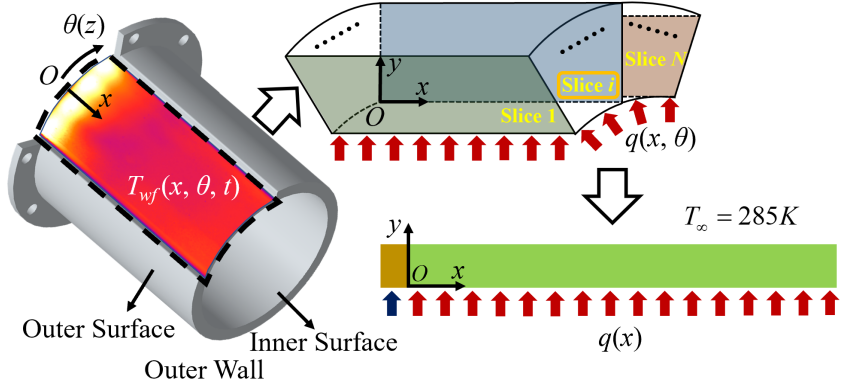

Figure 5. Schematic diagram of the physical model

In order to solve the temperature distribution $T_{wf}(x, t)$ at the upper boundary of the physical model when the heat flux $q_s(x)$ was given, the two-dimensional heat conduction equation without the internal heat source is used to calculate the heat conduction problem:

$$\rho c \frac{\partial T}{\partial t} = \frac{\partial}{\partial x}(\lambda \frac{\partial T}{\partial x}) + \frac{\partial}{\partial y}(\lambda \frac{\partial T}{\partial y}) \qquad (1)$$

where the physical properties $\rho$, $c$, and $\lambda$ are assumed as constant. Appendix A provides detailed discretization and solution settings for the heat conduction equation.

The process of solving the inverse problem of heat flux is presented in Figure 6. Initially, the heat flux is utilized in the heat conduction equation Eq.(1) to calculate the distribution of upper boundary temperature of the physical model. Then the distribution of heat flux is obtained by utilizing the inverse solving methods, and the error between the calculated and specified heat flux is assessed. Finally, the inverse solving methods are compared and examined for their accuracy in solving the heat flux distribution within RDCs.

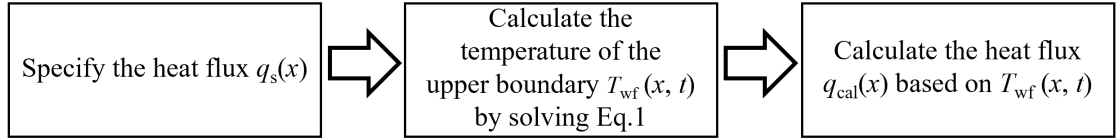

Figure 6. Research process of solving the inverse problem of heat flux

## 3.2 Heat flux calculation methods

### 3.2.1 Zero-dimensional method

The first heat flux calculation method is called the zero-dimensional (0D) method, which has been applied to the analysis of heat flux within various types of detonation combustors, such as the pulse detonation combustor[33] and the RDC[31]. According to the heat transfer theory, during the early stage after ignition, the temperature distribution of the combustor outer wall is mainly influenced by the initial conditions, and a stable temperature gradient has not been established yet, so the temperature of the combustor outer surface rises slowly, which is called the "non-regular regime". After that, the radial temperature gradient of the combustor outer wall becomes relatively stable, and the temperature of the combustor outer surface rises approximately linearly, which is called the "regular regime". The core theory of the 0D method is to calculate the heat flux utilizing the slope of the temperature change at the measuring point in the regular regime. The heat loss caused by natural convection heat exchange between the combustor outer wall and the environment is disregarded. The 0D method used in this study can be expressed as Eq.(2):

$$q_{cal}^{0D}(x) = c \cdot \rho \cdot \delta \cdot \frac{dT_{wf}(x,t)}{dt} \quad (2)$$

The expression $dT_{wf}(x, t)/dt$ represents the slope of the temperature change at the measurement point located at $x$ within the regular region.

### 3.2.2 L-M method

From Eq.(2), it is evident that the 0D method ignores the non-equilibrium heat conduction in the axial direction of the RDC. When a significant axial gradient of heat flux exits within the RDC, the calculated result may be a significantly deviate from the actual value. In comparison, the methods that calculate heat flux by solving the heat conduction problem can achieve higher accuracy theoretically. Problems solved through this type of method are commonly known as inverse heat conduction problems (IHCP). Scholars have conducted extensive research on this issue and have developed a series of solution methods such as the conjugate gradient method[34] and the Levenberg-Marquardt (L-M) method[35,36]. Among them, the L-M method has been widely applied in solving IHCP due to the advantages of fast convergence rate and high solution accuracy[37–40].

Heat flux on the inner surface of the outer wall is a boundary condition to be determined, which can be expressed using a series of estimated parameters and specific functional forms. The non-uniformly distributed heat flux $q_{\text{cal}}^{\text{LM}}$ is expressed as:

$$q_{\text{cal}}^{\text{LM}}(x) = \sum_{j=1}^{N} P_j C_j(x) \tag{3}$$

The parameters vector to be estimated is represented by $P_j$ in the equation, $N$ refers to the number of basic functions, and $C_j(x)$ denotes a specific type of function like sine, cosine et cetera. In this study, the cosine function is selected as the basic function due to its wider range of function values and stronger convergence during iterative solution[41]. Consequently, the Eq.(3) can be restated as follows:

$$q_{\text{cal}}^{\text{LM}}(x) = \sum_{j=1}^{N} P_j \cos[(j-1)\omega x] \tag{4}$$

where $\omega$ denotes the angular frequency, which can be expressed as $\omega = 2\pi/2L$. The L-M method is a gradient descent method widely recognized for its robustness and applicability to both linear and nonlinear inverse problems. The method is rooted in the principle of minimizing variance to evaluate the accuracy of the estimated parameter vector $\boldsymbol{P} = (P_j, j=1, 2, \ldots, N)^T$, and the variance is expressed as:

$$S(\boldsymbol{P}) = \sum_{i,j=1}^{m,n} \left[ T_{\text{wf},i}^j - T_{\text{cal},i}^j(\boldsymbol{P}) \right]^2 \tag{5}$$

where $T_{\text{wf},i}^j$ represents the measured temperature at location $x_i$ and time $j$, while $T_{\text{cal},i}^j(\boldsymbol{P})$ represents the calculated temperature at the same location and time based on the heat flux derived from the $\boldsymbol{P}$. $m$ represents the number of sampling points in $x$ direction, and $n$ represents the number of sampling times in $t$ direction.

The primary procedure of the L-M method is to iteratively determine the estimated parameter vector $\boldsymbol{P}$, which can be expressed as:

$$\boldsymbol{P}^{k+1} = \boldsymbol{P}^k + \left[ \left(\boldsymbol{J}^k\right)^T \boldsymbol{J}^k + \mu^k \boldsymbol{\Omega}^k \right]^{-1} \left(\boldsymbol{J}^k\right)^T \left[ \boldsymbol{T}_{\text{wf}} - \boldsymbol{T}_{\text{cal}}(\boldsymbol{P}^k) \right] \tag{6}$$

where $k$ represents the current iteration step of the calculation, and $\boldsymbol{J}^k$ is the sensitivity matrix, which is expressed as:

$$\boldsymbol{J}^k = \left( \frac{\partial \boldsymbol{T}_{\text{cal}}^T(\boldsymbol{P}^k)}{\partial \boldsymbol{P}^k} \right)^T \tag{7}$$

In Eq.(6), $\boldsymbol{\Omega}^k$ is a diagonal matrix defined as $\boldsymbol{\Omega}^k = \text{diag}[(\boldsymbol{J}^k) \cdot \boldsymbol{J}^k]$, and $\mu^k$ is a tuning factor whose value depends on the variance $S(\boldsymbol{P}^k)$ which is calculated during the iterative solving process. If $S(\boldsymbol{P}^k) \leq S(\boldsymbol{P}^{k-1})$, then $\mu^{k+1} = 0.1\mu^k$, if $S(\boldsymbol{P}^k) > S(\boldsymbol{P}^{k-1})$, then $\mu^{k+1} = 10\mu^k$. The calculation is considered to have converged when the following conditions are met:

$$\left|S\left(\boldsymbol{P}^{k+1}\right)-S\left(\boldsymbol{P}^{k}\right)\right|<\xi_{1}$$
$$\left|\left(\boldsymbol{J}^{k+1}\right)^{\mathrm{T}}\left[\boldsymbol{T}_{\mathrm{wf}}-\boldsymbol{T}_{\mathrm{cal}}\left(\boldsymbol{P}^{k+1}\right)\right]-\left(\boldsymbol{J}^{k}\right)^{\mathrm{T}}\left[\boldsymbol{T}_{\mathrm{wf}}-\boldsymbol{T}_{\mathrm{cal}}\left(\boldsymbol{P}^{k}\right)\right]\right|<\xi_{2} \quad (8)$$
$$\left|\boldsymbol{P}^{k+1}-\boldsymbol{P}^{k}\right|<\xi_{3}$$

where $\xi_1$, $\xi_2$ and $\xi_3$ represent the calculation convergence thresholds. The procedure of the L-M method is summarized in Figure 7.

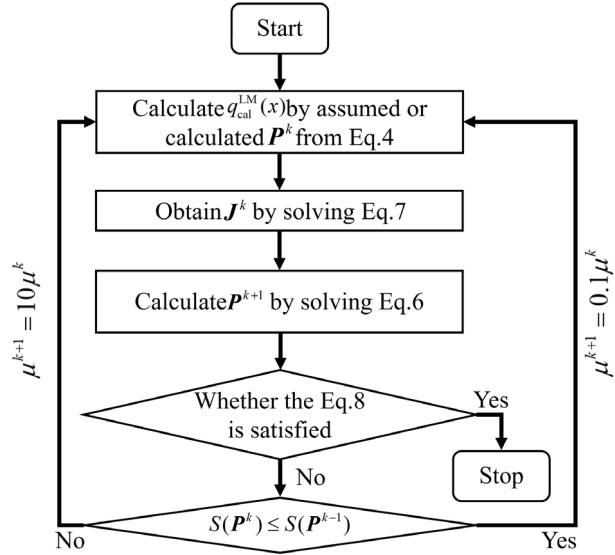

Figure 7. Calculation procedure of the L-M method

## 3.4 Comparison of heat flux calculation methods

### 3.4.1 Setup conditions

Based on the physical model illustrated in Figure 5, boundary conditions are established as follows: the left boundary ($x$ = -2 mm) corresponds to the end face of the locating slot, where exists contact thermal transfer. Considering that the dimension of interface thermal conductivity coefficient is the same as the dimension of convective heat transfer coefficient, this boundary is considered as convective heat exchange boundary. The right boundary ($x$ = 100 mm) corresponds to the end of the outer wall of the RDC. To minimize the impact of the flame on the infrared thermal imager, a baffle was placed at the end of the RDC during the experiments. This boundary is assumed to be adiabatic by neglecting the heat exchange between the baffle and the outer wall. The top boundary ($y$ = 3 mm) corresponds to the combustor outer surface, and this boundary is treated as the convective heat exchange boundary. The heat flux at the lower boundary ($y$ = 0 mm) is set as $q_{\mathrm{spec}}(x)$, where $q_{\mathrm{spec}}(x < 0) = 0$ MW/m² is fixed. The specific setting method of $q_{\mathrm{spec}}(x > 0)$ is as follows.

In order to approximate the heat flux distribution within RDCs[22,24,30], nine heat flux distribution are defined, as shown in Figure 8, in conjunction with the experimental results of the temperature field. Each distribution is represented by a polynomial equation, as shown in Eq.(9). The coefficients in the equation are calculated based on the heat flux $q_{spec}|_{x_A, x_B, x_C}$ and its gradient $\frac{dq_{spec}}{dx}|_{x_A, x_B, x_C}$ at fixed-points A, B and C. The fixed-point values for different heat flux distributions are presented in Table 1. The nine types of heat flux exhibit a consistent distribution pattern, characterized by higher heat flux and larger gradient in the region of $0 < x < 40$ mm, and lower heat flux density and smaller gradient in the remaining regions. This distribution pattern aligns with the observed characteristics of the temperature field observed in the experiment.

$$q_{spec}(x) = \begin{cases} 0.4x^4 + a_3^1 x^3 + a_2^1 x^2 + a_1^1 x + a_0^1, & x < 40 \ mm \\ 0.01x^4 + a_3^1 x^3 + a_2^1 x^2 + a_1^1 x + a_0^1, & x \geq 40 \ mm \end{cases} \quad (9)$$

Table 1. The fixed-point values for different heat flux distributions

| Case | $q_{spec}$ (MW/m²) [A, B, C] | $dq_{spec}/dx$ (MW/m³) [A, B, C] |
|---|---|---|
| 1.1 |  | [0, 0, 0] |
| 1.2 | [1.25, 0.60, 0.65] | [0.02, 0, 0] |
| 1.3 |  | [0.06, 0, 0] |
| 2.1 |  | [0, 0, 0] |
| 2.2 | [1.50, 0.60, 0.65] | [0.02, 0, 0] |
| 2.3 |  | [0.06, 0, 0] |
| 3.1 |  | [0, 0, 0] |
| 3.2 | [1.75, 0.60, 0.65] | [0.02, 0, 0] |
| 3.3 |  | [0.06, 0, 0] |

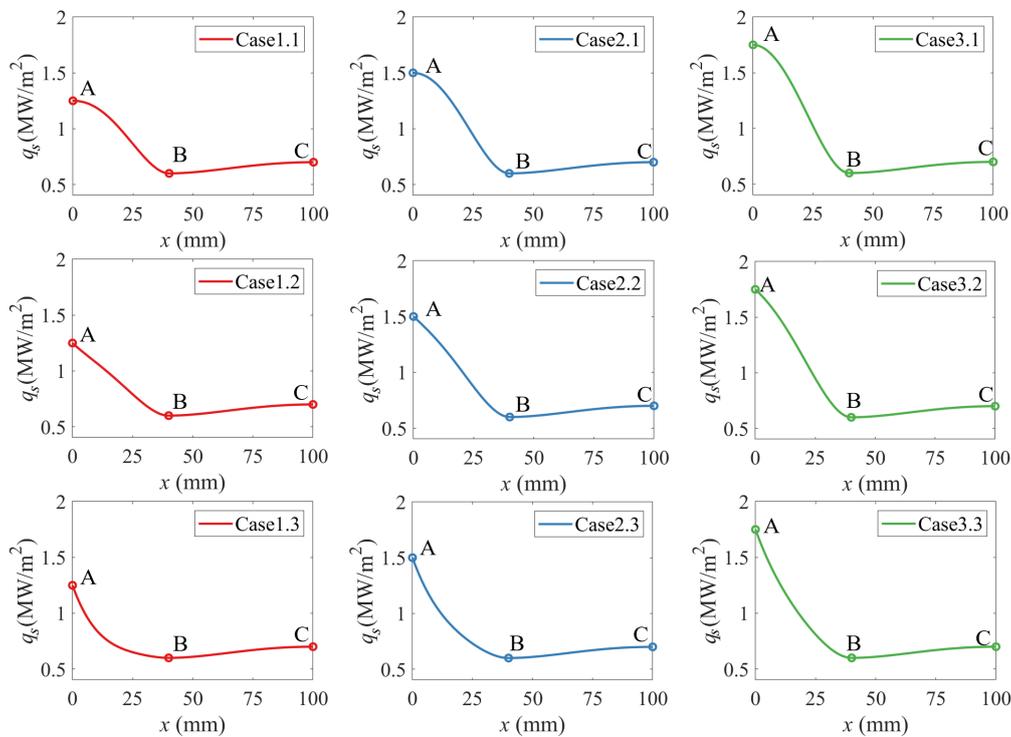

Figure 8. Heat flux distribution at the lower boundary of the physical model

The outer wall of the RDC has the constant specific heat capacity $c = 500$ J/(kg·K) and the constant density $\rho = 8000$ kg/m³. The moment of applying the provided heat flux to the physical model is set as the zero-time point ($t_0 = 0$ ms). The length of the physical model in $x$ direction $x_L$ is set to 100 mm, and the length in $y$ direction $y_L$ is set to 3 mm, with the spatial step size $\Delta_x = 0.2$ mm. The total calculation time $t_{total}$ is set to 1000 ms, with the time step size $\Delta_t = 0.02$ s. The temperature distribution $T_{wf}(x, t)$ on the top boundary of the physical model is recorded at various locations and times. The 0D method and the L-M method are used to inversely calculate the heat flux distribution along the axial direction of the RDC, and the results are compared to the $q_{spec}(x)$.

### 3.4.2 Numerical simulation results

The results comparing the heat flux calculated by the 0D method $q_{cal}^{0D}(x)$ and the L-M method $q_{cal}^{LM}(x)$ ($N = 15$) with the specified heat flux $q_{spec}(x)$ are presented in Figure 9. The maximum and minimum values of the heat flux are labeled with triangular and circular points. It can be seen that there is a certain difference between the $q_{cal}^{0D}(x)$ and $q_{spec}(x)$, especially in the region with $x < 20$ mm. In this region, $q_{cal}^{0D}(x)$ exhibits an upward trend followed by a downward trend, whereas $q_{spec}(x)$ shows a continuous decrease. Additionally, there is a significant deviation in both the maximum values and their corresponding positions. The $x$ direction heat conduction is obvious in the region with large heat flux gradient, which cause a substantial deviation between the heat flux calculated by the 0D method $q_{cal}^{0D}(x)$ and specified heat flux $q_{spec}(x)$. Compared to the results obtained by 0D method, the difference between $q_{cal}^{LM}(x)$ and $q_{spec}(x)$ is smaller in terms of the maximum value and the heat flux distribution trend.

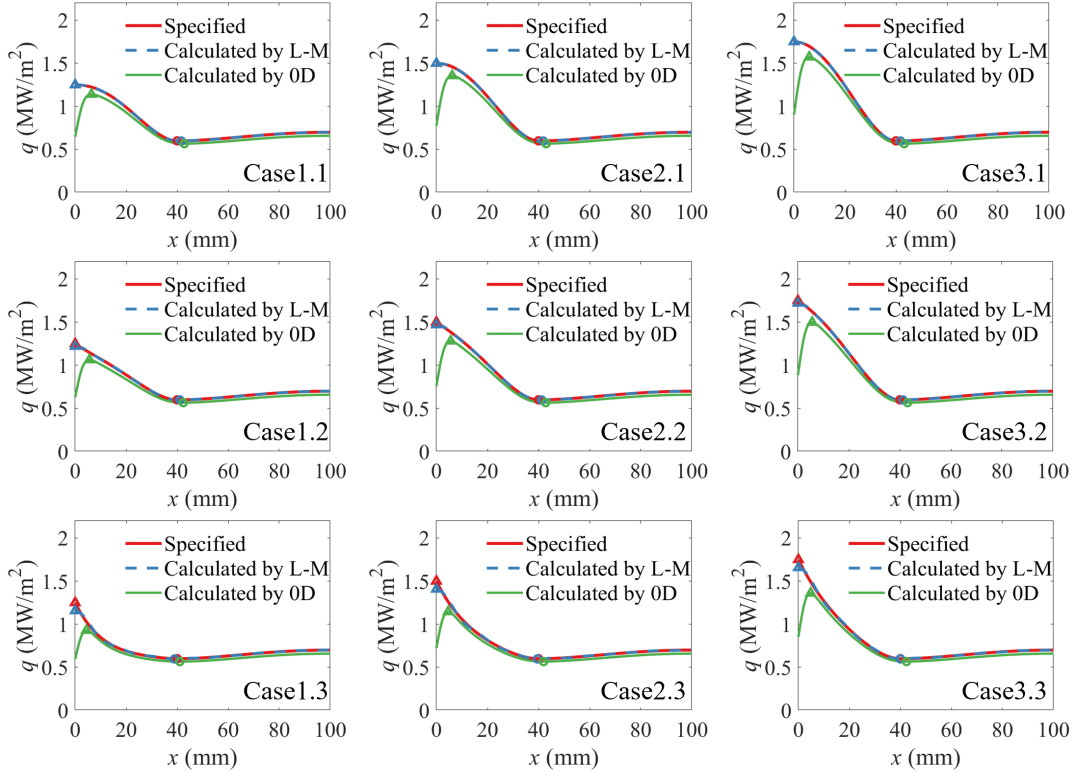

Figure 9. The comparison between the calculated heat flux ($q_{cal}^{LM}(x)$, $q_{cal}^{LM}(x)$) and the specified heat flux $q_{spec}(x)$.

Figure 10 provides an overview of the relative error of the maximum values and their positions between calculation results ($q_{cal}^{0D}(x)$, $q_{cal}^{LM}(x)$) and $q_{spec}(x)$ for different forms of heat flux. It can be seen that the relative error of calculation is more closely related to the heat flux spatial gradient at the maximum value than to the maximum value itself, for both 0D method and L-M method. The results indicate that the relative calculation errors of L-M method are significantly smaller than the relative calculation errors of 0D method. The analysis above indicate that the L-M method can obtain accurate results for the current physical model and heat flux distribution pattern.

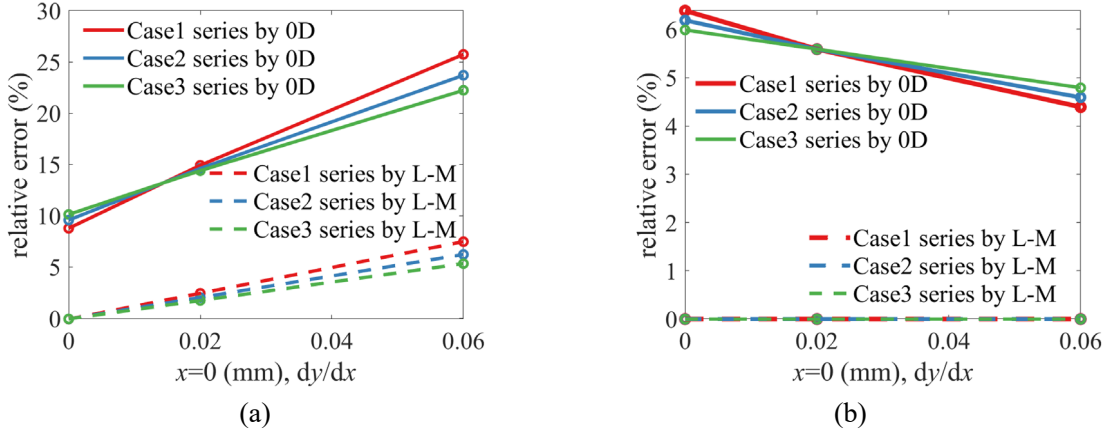

Figure 10. Relative errors of calculation results: a) relative error of maximum heat flux, b) relative error of maximum heat flux position.

The impact of the number of basic functions on $q_{cal}^{LM}(x)$ is demonstrated using case 1.3 as an example. Figure 11 displays the comparison between $q_{cal}^{LM}(x)$ for different number of basic functions and $q_s(x)$. Figure 12 depicts the influence of the number of basic functions on the relative error of calculated results. The results demonstrate that increasing the number of basic functions results in a reduction in the relative error of the maximum heat flux. The L-M method employs a combination of estimated parameters and specific basis functions to approximate the distribution form of the target physical quantity. Increasing the number of basic functions is beneficial for improving the solution accuracy of the L-M method, but also increases the calculation workload. Based on the current physical model and heat flux distribution, and considering the results depicted in Figure 11 and Figure 12, the number of basic functions $N$ was set to 15 for subsequent calculations.

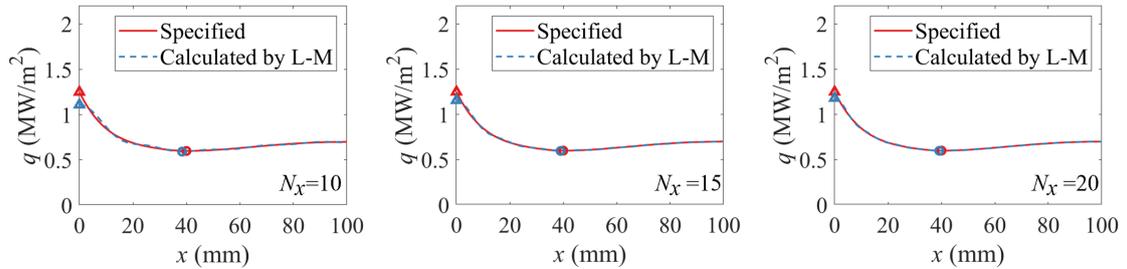

Figure 11. Comparison of heat flux calculated by the L-M method $q_{cal}^{LM}(x)$ using varying numbers of basic functions and specified heat flux $q_{spec}(x)$.

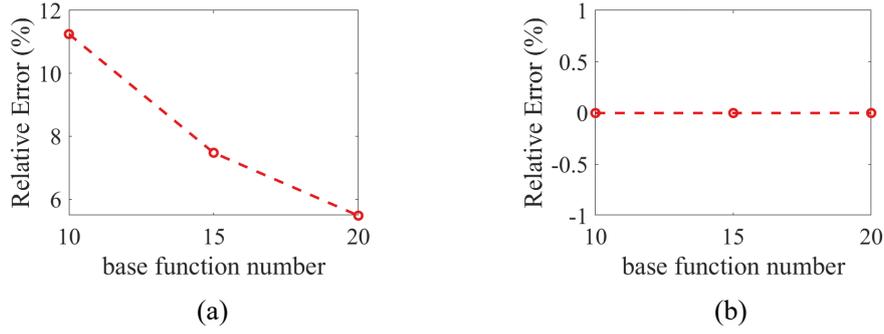

Figure 12. The influence of the number of basic functions on the relative error of results calculated by the L-M method: a) relative error of maximum heat flux, b) relative error of maximum heat flux position

## 4. Experimental measurements and comprehensive analysis

### 4.1 Data analysis in experiments

The thermocouple was used to calibrate the temperature data obtained from the infrared thermal imager. A thermal response model of the thermocouple is established, and the theoretical analysis of measurement error is conducted with consideration of the thermocouple performance parameters. The detailed process is described in Appendix B. The results show that within a certain range of error, the film K-type thermocouple used in the experiment can reflect the temperature changes of the outer wall of RDC. The temperature data obtained by the thermocouple and the infrared thermal imager at the same measurement point are compared, as shown in Figure 13. During the initial stage from ignition, the infrared thermal imaging data was distorted due to vibration and the influence of flames, but the data returned to normal about 500 ms after ignition. It can be seen that after 500 ms from ignition, the relative error between the temperature data obtained by the infrared thermal imager and the thermocouple remains below 2%. The error primarily originates from the vibration occurring during RDC operation and the deviation of the emissivity of the temperature measurement area coating.

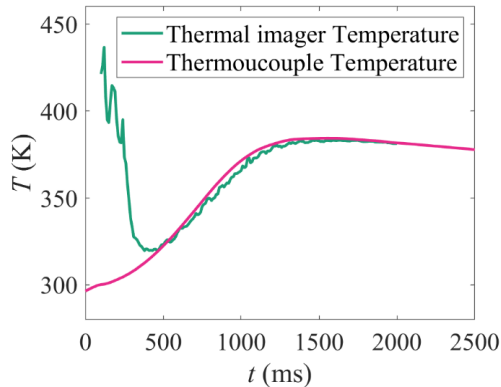

Figure 13. Comparison between the temperature measured by the infrared thermal imager and the temperature measured by a thermocouple.

Figure 14 depicts the image captured by the infrared thermal imager and the temperature field distribution in the measurement area at 800 ms after ignition under the condition of $\dot{m}_O$ = 565 g/s and ER = 0.64. The contour lines of the temperature field appear to be distorted and irregular due to the minor variations in the thickness of the sprayed coating in the measurement area, boundary heat conduction due to different wall thicknesses, and high-frequency vibrations during RDC operation. It can be observed that the temperature of the outer wall varied along the axial ($x$) direction and circumferential ($z$) direction, revealing four local high-temperature areas in the head of the measuring area that coincide with the nozzle position. Figure 15 displays the temperature distribution curve of the outer wall along the $x$ and $z$ directions in the measuring area. It can be seen from Figure 15(a) that in the upstream region of the RDC where $0 \leq x \leq 20$ mm, the temperature and temperature gradient are higher, whereas in the downstream region of the RDC where $x > 40$ mm, the temperature and temperature gradient are lower. Figure 15(b) indicates that the temperature distribution along the $z$ direction of the outer wall is inconsistent at different $x$ positions. The outer wall temperature along the $z$ direction is notably non-uniform in the upstream region of the RDC, with a relative difference of approximately 6% between the maximum and minimum temperature when taking $x = 5$ mm as an example. Conversely, the outer wall temperature along the $z$ direction is almost evenly distributed in the downstream of the RDC when taking $x = 50$ mm as an example.

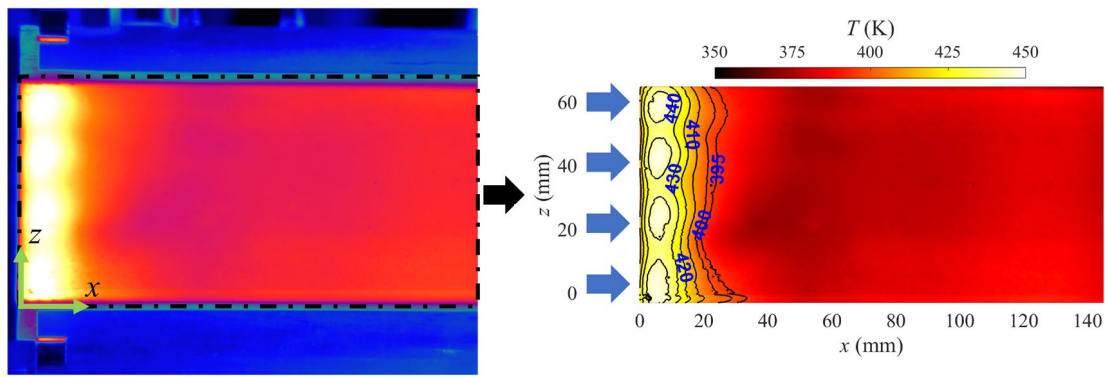

Figure 14. The image captured by the infrared thermal imager and the temperature field distribution, with the position of the injector marked by blue arrows ($\dot{m}_o$=565g/s, ER=0.64).

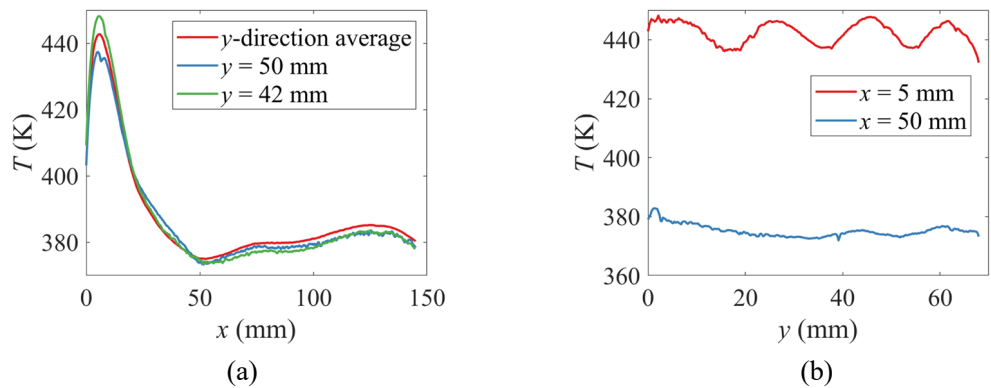

(a)      (b)

Figure 15. Temperature distribution of the outer wall of the RDC: a) temperature distribution in the $x$ direction, b) temperature distribution in the $z$ direction.

Figure 16 shows the typical sensor signals obtained during experiment. It can be seen that the mass flow of both oxidizer and kerosene remained relatively stable during RDC operation, with relative errors change remained below 5%. By analyzing the time interval between adjacent detonation wave pressure peaks using the 50 ms data from the high-frequency pressure sensor, the relationship between the propagation speed of the detonation wave and time is analyzed. It can be seen that the detonation wave propagation speed was relatively stable during experiment, indicating that a stable propagating detonation wave was achieved and the combustion mode did not change in the experiment.

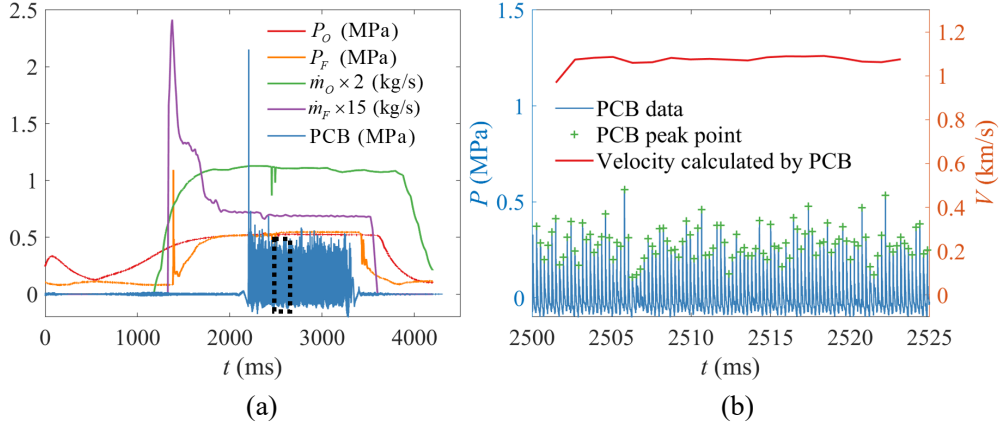

Figure 16. Typical sensor signal in the experiment (a) and analysis of the high-frequency pressure data (b).

## 4.2 Demonstration of wall heat fluxes

Figure 17 shows the heat flux in the temperature measurement zone of the RDC under $\dot{m}_o$=565g/s and different ER conditions using heat flux calculation methods discussed in the previous sections. According to the high-frequency pressure data, the combustion mode in RDC was stable under three ER conditions, so the heat flux distribution could be considered as steady state. The heat flux distribution in both the axial and the circumferential directions of the RDC was non-uniform. Specifically, the axial non-uniformity is evidenced by the peak value of heat flux appearing in the head area of the RDC where $x \leq 10$ mm, with relatively lower heat flux values in the middle and downstream regions of the RDC. The circumferential non-uniformity primarily manifests as periodical high heat flux areas in the head region of the RDC, specifically for $x \leq 10$ mm. These high heat flux areas align with the position of the injector. The presence of recirculation zones near the injectors is speculated to be a possible reason for the formation of the high heat flux area. The similar flow structures have also been reported in other coaxial injection combustors[42]. The high temperature combustion products entering the recirculation zone after the detonation wave cannot be discharged in time, which results the local increase of heat flux.

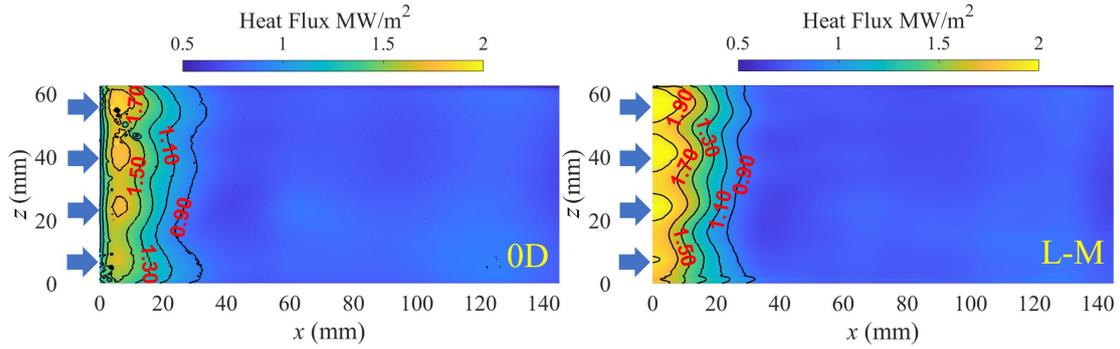

(a) ER = 0.44

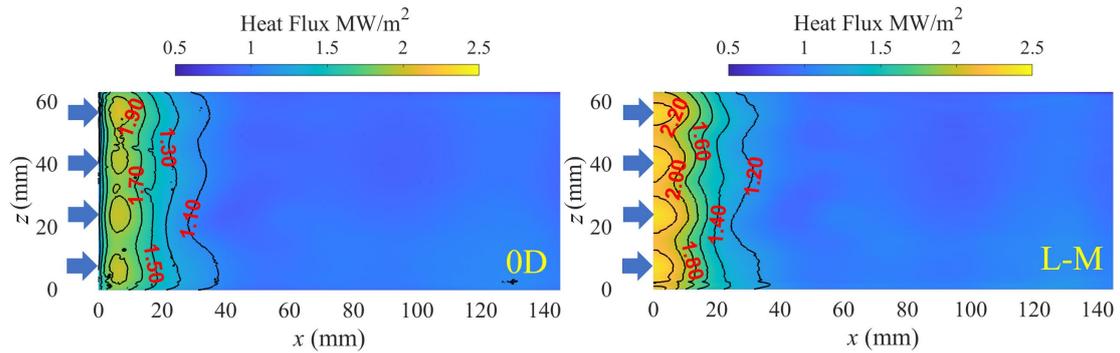

(b) ER = 0.55

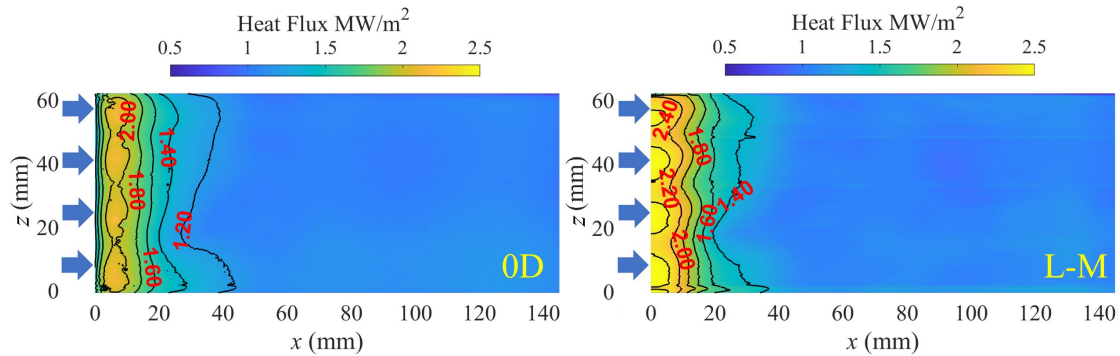

(c) ER = 0.64

Figure 17. The distribution of heat flux in the temperature measurement area of RDC obtained by two kinds of heat flux calculation methods under different ER conditions, with the position of the injector marked by blue arrows.

Figure 18 presents a comparison of the axial distribution of the circumferential average heat flux calculated using both the 0D and L-M methods in the temperature measurement area of the RDC. The results clearly demonstrate that there is minimal difference in the heat flux calculated by the two methods in the downstream region of the RDC. However, significant disparities are observed in the heat flux calculated by two methods in the upstream region of the RDC, specifically for $x \leq 20$ mm. The aforementioned results corroborate the findings obtained from the previous analysis using the heat conduction equation. Some existing researches have provided experimental evidence indicating a significant axial heat flux gradient in the upstream region of the RDC[31]. Considering the geometric configuration of the experimental device and the results obtain in this study, it is highly probable that the experimental results in existing research are attributed to the boundary heat conduction of the outer wall. As a result, the actual heat flux distribution may deviate from the calculated results.

Based on the results calculated by the L-M method, it is noticeable that the peak heat flux rises as the ER increases at the current oxidizer mass flow conditions. Integrating the circumferential average heat flux under different conditions along the $x$ direction, and the integral result is multiplied by the circumference of the outer wall of the RDC to represent the thermal heat rate of the RDC. The relationship between the heat rate of the RDC and the ER is shown in Figure 19, which indicates that the heat rate of the RDC increases with the rise of the ER.

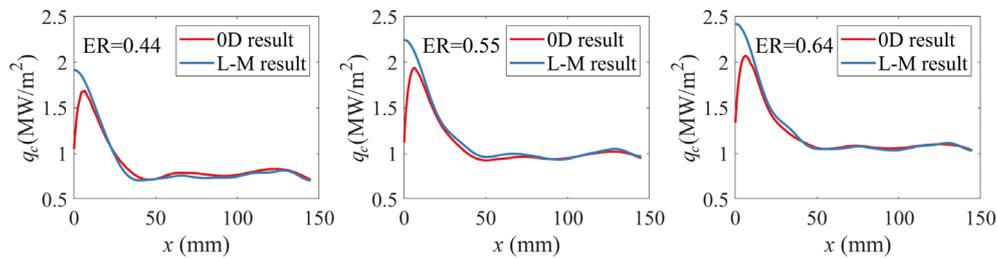

Figure 18. Comparison of the axial distribution of circumferential average heat flux calculated by the 0D and L-M methods in the temperature measurement area.

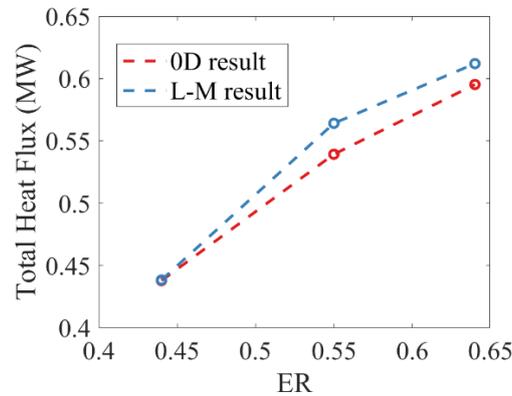

Figure 19. The relation between the thermal heat rate within RDC and ER.

## 5. Conclusion

In this study, the comprehensive analysis method has been established to obtain the non-uniform wall heat flux distribution in the RDC based on the measurements of temperature of combustor outer surface by the high-speed infrared thermal imager. The calculation accuracy and applicability of the L-M method is verified based on the theoretical model of the thermal conduction calculation. Furthermore, the distribution of wall heat flux of RDC can be comparatively analyzed based on the above method. The conclusions of this study are as follows:

The analysis of the calculation results under steady-state non-uniform heat flux condition demonstrates that the L-M method, based on solving IHCP, is capable for the cases existing significant heat flux spatial gradients with the consideration of the influence of axial heat conduction, compared to the 0D method. Together with the measurements of temperature of combustor outer surface, the wall heat fluxes can be numerically obtained for the present RDC. The heat flux and its gradient are larger in the upstream region, compared to the downstream region of the RDC. Under the same mass flow rate of oxidizer, it is observed that both the peak heat flux and the heat rate of the RDC increase with the equivalence ratio rises within the lean combustion range.

Further investigations will be carried out using the measurement and calculation method introduced in this study to analyze the influencing factors of heat flux distribution in experiments with wide range of mass flow rate and equivalence ratio.

# 6. Appendix

## A: Numerical procedure setting

In order to solve the heat conduction problem, the physical model needs to be discretized into several subregions and the nodes in each subregion need to be determined. There are two commonly used methods for setting nodes in heat conduction calculation: one is the external node method (also known as the element vertex method), which assumes that the node is located at the corner of the subregion, and the other is the internal node method (also known as the element center method), which assumes that the node is located at the center of the subregion[43]. The internal node method is used in this study, shown in Figure 20, and the temperature distribution in every subregion is considered uniform. Eq.(1) can be represented in the implicit discretized form:

$$a_P T_P = a_E T_E + a_w T_w + a_N T_N + a_S T_S + b \tag{10}$$

The node coefficient can be expressed as:

$$a_E = \frac{\Delta y}{(\delta x)_e / \lambda_e}, \ a_W = \frac{\Delta y}{(\delta x)_w / \lambda_w}, \ a_N = \frac{\Delta x}{(\delta x)_n / \lambda_n}, \ a_S = \frac{\Delta x}{(\delta x)_s / \lambda_s},$$

$$a_P = a_E + a_W + a_N + a_S + a_P^0 - S_P \Delta x \Delta y,$$

$$a_P^0 = \frac{(\rho c)_P \Delta x \Delta y}{\Delta t}, \quad b = S_C \Delta x \Delta y + a_P^0 T_P^0.$$

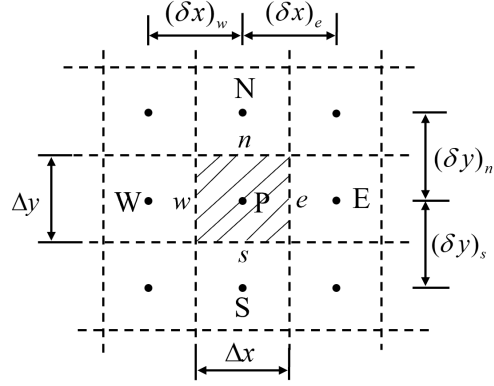

Figure 20. internal node method for the physical model in this study

Boundary conditions are handled using the additional source term method. This approach regards the heat entering or leaving the boundary as a source term in the adjacent subregion[43]. The four boundaries shown in Figure 5 can be treated as follows:

Lower boundary:

$$\begin{aligned} b &= q_s \cdot dx, \\ a_S &= 0, \\ a_P &= a_P^0 + a_E + a_W + a_N + a_S \end{aligned} \quad (11)$$

Left boundary:

$$\begin{aligned} b &= \frac{T_\infty \cdot dy}{1/h_c + dx/(2\cdot\lambda)}, \\ a_E &= 0, \\ a_P &= a_P^0 + a_E + a_W + a_N + a_S + \frac{T_\infty \cdot dy}{1/h_c + dx/(2\cdot\lambda)} \end{aligned} \quad (12)$$

In Eq.(12), $h_c$ represents the contact thermal conductivity coefficient, for stainless steel contact surfaces with normal surface roughness and medium contact pressure(1-10 atm), the range of contact thermal conductivity coefficient is between 2000-3700 W/(m²·K), $h_c$ is set to 3000 W/(m²·K) in this study.

Upper boundary:

$$\begin{aligned} b &= \frac{T_\infty \cdot dx}{1/h + dy/(2\cdot\lambda)}, \\ a_N &= 0, \\ a_P &= a_P^0 + a_E + a_W + a_N + a_S + \frac{dx}{1/h + dy/(2\cdot\lambda)} \end{aligned} \quad (13)$$

In Eq.(13), $T_\infty$ is environment temperature which set to 285 K, $h$ is the convective heat transfer coefficient between the outer wall and environment, for normal temperature air, $h$ is typically in the range of 3-10 W/(m²·K), while a simplified value of 10 W/(m²·K) is utilized in this study.

Right boundary:

$$b = 0,$$
$$a_W = 0, \tag{14}$$
$$a_P = a_P^0 + a_E + a_W + a_N + a_S$$

Using the additional source term method to handle boundary nodes allows them to be iterated together with internal nodes. The Gauss-Seidel (G-S) iterative method was used to solve the Eq.(10) at each time step. The iterative method is an approximate solution method for linear equation systems that is easy to program. The solution accuracy is determined by the number of iterations, with higher number of iterations resulting in higher accuracy. The calculation convergence threshold is set to 1e-7 in this study considering the calculation time and accuracy. When the temperature difference at the same subregion calculated by two adjacent iterations is less than the calculation convergence threshold, the iteration is stopped.

## B: Thermocouple model and measurement error analysis

Thermocouples detect temperature by measuring the electromotive force generated by the temperature difference between the measurement end (hot end) and the cold end (compensating end). The response lag time, which is influenced by the heat capacity of the measuring end, is a crucial parameter for thermocouples. A shorter response lag time of the thermocouple leads to higher accuracy in temperature measurement. To assess the disparities between the temperature data obtained from the thermocouple used in this study and the actual temperature at the measurement point, the temperature output data of the thermocouple is interpreted based on heat transfer theory:

$$h_p(T_r^i - T_c^i)\Delta t = C \cdot m_c \cdot (T_c^{i+1} - T_c^i) \tag{15}$$

where $h_p$ is the thermal conductivity at the contact interface between the thermocouple measurement end and the surface of the object being measured, $T_r^i$ represents the actual temperature at the measurement point at time $i$, $T_c^i$ and $T_c^{i+1}$ represent the temperature at the measurement end of the thermocouple at time $i$ and time $i+1$, $C$ denotes the specific heat of the thermocouple measurement end material, and $m_c$ represents the mass of the measurement end. Eq.(15) can be expressed in differential form as follows:

$$[T_r(t) - T_c(t)] = \frac{C \cdot m_c}{h_p} \cdot \frac{dT_c}{dt} \tag{16}$$

$\frac{C \cdot m_c}{h_p}$ can be replaced by $\eta$ which represents the response lag characteristics of the thermocouple. If $T_r$ is constant, the ordinary differential equations shown as Eq.(16) can be solved as:

$$\frac{T_c(t) - T_r}{T_c(0) - T_r} = e^{\frac{-t}{\eta}} \quad (17)$$

In the previous sections, it is mentioned that the response time of the thermocouple used in the experiment to track 95% of the instantaneous temperature change of the measuring point is 0.08 s. By substituting this value into Eq.(17), the value of $\eta$ can be calculated as 0.0267. It is observed that the temperature of the measurement point increased approximately linearly after some time after ignition, Eq.(16) can be changed as:

$$k \cdot t = \eta \frac{dT_c}{dt} + T_c(t) \quad (18)$$

$k$ represents the rate of the temperature change, which is set to 100 K/s. Eq.(18) is solved and the comparison between the temperature of the measurement point $T_r(t)$ and the temperature of the thermocouple measurement end $T_c(t)$ is shown in Figure 21(a). A discrepancy $\Delta T$ exists between $T_r(t)$ and $T_c(t)$, which illustrates the influence of the response lag characteristic of the thermocouple on temperature measurement. Figure 21(b) shows the relation between $\Delta T$ and time $t$, it can be seen that $\Delta T$ reaches stability at 0.16 s. As the value of $k$ changes, it is observed that the temperature deviation increases. However, the time required for the temperature deviation to stabilize remains constant. Considering the working time of RDC, it can be assumed with a certain margin of error that the temperature output of the thermocouple represents the actual temperature of the measurement point.

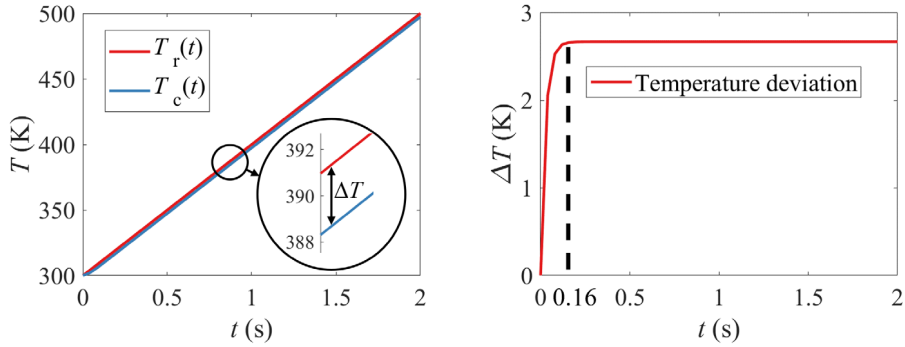

Figure 21. The comparison between the temperature of measuring point $T_r(t)$ and the temperature of the thermocouple measuring end $T_c(t)$ (left) and the discrepancy between $T_r(t)$ and $T_c(t)$ (right).